\documentclass[twocolumn]{aastex631}

\usepackage{amsmath}

\def\arcdeg{$^\circ$}

\def\kmskpc{\,km\,s$^{-1}$\,kpc$^{-1}$}
\def\H2{$\rm H_2$}

\def\EffPotL1{$\Phi_{\rm eff,b}(R_{\rm L1/L2})$}

\begin{document}

\title{
Bar-driven Streaming Motions Mimic a Massive Bulge in the Inner Milky Way
}

\email{babajn2000@gmail.com; junichi.baba@sci.kagoshima-u.ac.jp; jun.baba@nao.ac.jp}

\author[0000-0002-2154-8740]{Junichi Baba}
\affiliation{Amanogawa Galaxy Astronomy Research Center, Kagoshima University, 1--21--35 Korimoto, Kagoshima 890-0065, Japan.}
\affiliation{National Astronomical Observatory of Japan, Mitaka, Tokyo 181-8588, Japan.}

\begin{abstract}
The circular speed curve of the Milky Way provides a key constraint on its mass distribution, reflecting the axisymmetric component of the gravitational potential. This is especially critical in the inner Galaxy ($R \lesssim 4$~kpc), where non-axisymmetric structures such as the stellar bar and nuclear stellar disk strongly influence dynamics. However, significant discrepancies remain between circular speed curves inferred from stellar dynamical modeling and those derived from the terminal-velocity method applied to gas kinematics. To investigate this, we perform three-dimensional hydrodynamic simulations including cooling, heating, star formation, and feedback, under a realistic gravitational potential derived from stellar dynamical models calibrated to observational data. This potential includes the Galactic bar, stellar disks, dark matter halo, nuclear stellar disk, and nuclear star cluster. We generate synthetic longitude–velocity ($l$–$v$) diagrams and apply the terminal-velocity method to derive circular speeds. The simulated gas reproduces the observed terminal-velocity envelope, including a steep inner rise. We find this feature arises from bar-driven non-circular motions, which cause the terminal-velocity method to overestimate circular speeds by up to a factor of 2 at $R \sim 0.4$~kpc, and enclosed mass by up to a factor of 4. These results suggest that inner gas-based rotation curves can significantly overestimate central mass concentrations. 
The steep inner rise in gas-derived circular speeds does not require a massive classical bulge but can be explained by bar-induced streaming motions. 
Rather than proposing a new mechanism, our study provides a clear, Milky Way-specific demonstration of this effect, emphasizing the importance of dynamical modeling that explicitly includes non-circular motions for accurate mass inference in the inner Milky Way.
\end{abstract}


\section{Introduction}
\label{sec:Intro}

\begin{figure*}
\begin{center}
\includegraphics[width=0.9\textwidth]{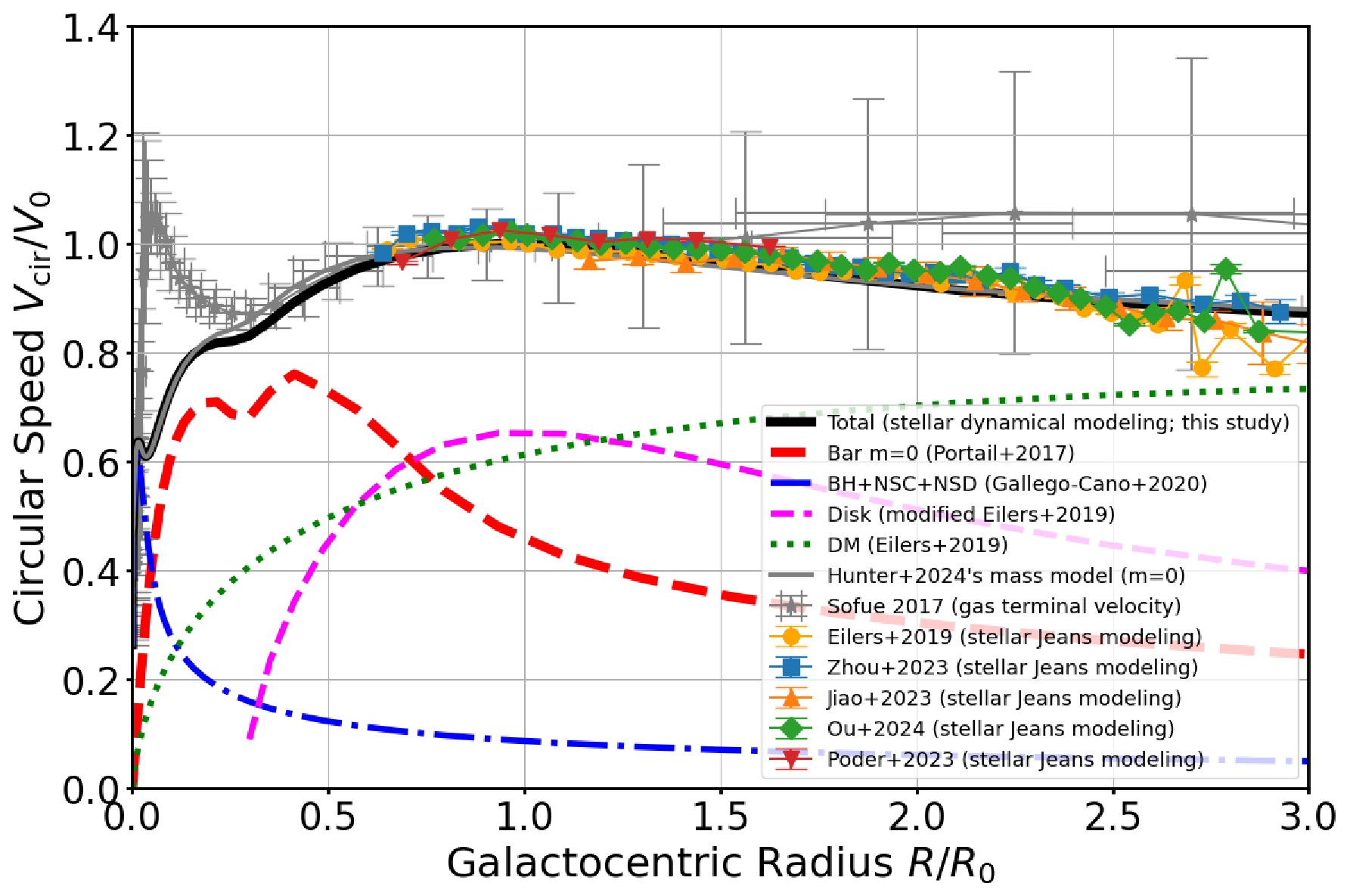}
\end{center}
\caption{
Circular speed curves in the Milky Way from different tracers and modeling methods, normalized by the local standard of rest ($R_0$, $V_0$). The gray curve with star markers shows the gas-based circular speed from the terminal-velocity method \citep{Sofue2017}. The thick black line represents the circular speed computed from the axisymmetric ($m = 0$) component of a barred potential based on stellar dynamical modeling. Contributions from individual components are shown: bar ($m=0$, red dashed; \citealt{Portail+2017, Sormani+2022agama}), stellar disk (magenta dashed; \citealt{Eilers+2019}), dark matter halo (green dotted; \citealt{Eilers+2019}), and central mass (NSD and NSC, blue dot-dashed; \citealt{Gallego-Cano+2020, Chatzopoulos+2015,Sormani+2020nsd}). 
The thin gray solid line shows the circular speed computed from the $m=0$ component of the mass model used in \citet{Hunter+2024}.
Data points with error bars show circular speeds inferred from axisymmetric Jeans modeling using \textit{Gaia} and spectroscopic data \citep{Eilers+2019, Zhou+2023, Jiao+2023, Ou+2024, Poder+2023}.
In the inner region ($R/R_0 \lesssim 0.3$), the gas-based curve rises more steeply than the stellar-based ones.
This study focuses on this region, where the terminal-velocity method is expected to be most affected by non-circular motions.
}
\label{fig:rc_comp}
\end{figure*}

Understanding the mass distribution of the Milky Way is a fundamental goal in Galactic astronomy \citep[][for a review]{Courteau+2014}. One of the most direct observational constraints on this distribution comes from the circular speed curve (or rotation curve)\footnote{
The term ``rotation curve'' is commonly used in the literature to describe the radial profile of rotational velocities of stars and gas within galaxies. However, it can be ambiguous, particularly in regions with strong non-axisymmetric structures, where the azimuthally averaged rotational velocity (i.e., mean $V_\phi$) may deviate significantly from the circular speed defined by centrifugal equilibrium. In this study, we prefer the term ``circular speed curve'' to explicitly refer to the velocity that balances centrifugal force with the ``axisymmetric component'' of the gravitational potential. Nonetheless, we occasionally use the term ``rotation curve'' when the meaning is clear from context.
}, which reflects the axisymmetric component of the gravitational potential \citep[][for a review]{Sofue2017}. In the inner Milky Way ($R \lesssim 4~\mathrm{kpc}$), accurately recovering the circular speed curve is particularly important, as the dynamics are significantly influenced by non-axisymmetric structures such as the central bar and spiral arms. Observations from \textit{Gaia} and large-scale spectroscopic surveys have greatly improved our understanding of the stellar disk and bulge, revealing a boxy/peanut-shaped bulge that smoothly transitions into a long bar extending to $\sim$5~kpc \citep[][]{WeggGerhard2013,Portail+2017}, inclined by $\phi_{\rm bar}\sim$25$^\circ$ relative to the Sun-Galactic center line. The bar's pattern speed is estimated to be $\sim$35-40~\kmskpc{} \citep[][for a review]{HuntVasiliev2025}.

For the stellar component, circular speed curves in the outer disk have been derived by fitting axisymmetric Jeans models to the phase-space distribution of red giant branch (RGB) stars \citep[][]{Eilers+2019,Zhou+2023,Jiao+2023,Ou+2024,Poder+2023}. These models, however, are not applicable to the central few kpc, where the assumption of axisymmetry breaks down and \textit{Gaia}-quality data are scarce. In contrast, made-to-measure (M2M) $N$-body modeling \citep[][]{SeyerTremaine1996} has enabled the reconstruction of the three-dimensional stellar mass distribution in the bar and bulge using ground-based photometric and spectroscopic surveys \citep[][]{Portail+2017}. The nuclear region ($R \lesssim 300~\mathrm{pc}$), including the nuclear stellar disk (NSD) and nuclear star cluster (NSC), has been modeled with stellar dynamical techniques based on near-infrared observations \citep[][]{Launhardt+2002,Chatzopoulos+2015,Gallego-Cano+2020,Sormani+2020nsd,Sormani+2022nsd}.

For the gas component, the most widely used method in the inner Milky Way is the ``terminal-velocity'' (or ``tangent-point'', TP) method, which estimates the circular speed from the maximum observed line-of-sight (LOS) velocity along a given Galactic longitude, assuming purely circular motion \citep[][]{BurtonGordon1978,Clemens1985,Sofue2013,Sofue2017}. 
While this method has been instrumental in constructing rotation curves from CO and H\,I emission, it neglects non-circular streaming motions induced by the bar, which were first pointed out in early studies of Galactic gas dynamics \citep[][]{LisztBurton1980,MulderLiem1986,BlitzSpergel1991,Binney+1991} and later explored in theoretical and numerical models \citep[e.g.,][]{Fux1999,EnglmaierGerhard1999,WeinerSellwood1999,Rodriguez-FernandezCombes2008,Baba+2010,Pettitt+2014,Sormani+2015a,Hunter+2024,Dura-Camacho+2024}. These non-circular motions can lead to biased mass estimates if not properly accounted for \citep[][]{KodaWada2002,Chemin+2015}.

Recent advances in stellar dynamical modeling, particularly through M2M techniques, have enabled increasingly accurate reconstructions of the mass distribution in the inner Milky Way. Figure~\ref{fig:rc_comp} compares representative rotation curves derived from different tracers and methods: gas kinematics based on the terminal-velocity method \citep[][]{Sofue2017}\footnote{Data available at \url{https://www.ioa.s.u-tokyo.ac.jp/~sofue/htdocs/2017paReview/}}, and stellar dynamical models calibrated to observed stellar data \citep[][]{Portail+2017,Sormani+2022agama,Eilers+2019}. In the bar region ($R/R_0 \lesssim 0.4$), stellar models yield a moderately rising rotation curve, whereas the gas-based curve rises much more steeply.
If interpreted at face value, this steep inner rise in the gas-based curve would imply the presence of a substantial mass concentration within the central few kiloparsecs. Such a feature has been previously interpreted as evidence for either a massive classical bulge \citep[][]{Sofue+2009,Sofue2013} or a compact dark matter component. However, stellar kinematic data in the Galacti bulge region suggest otherwise: studies such as \citet{Shen+2010} have argued against the presence of a massive classical bulge, showing instead that the observed kinematics are consistent with a bar-dominated pseudo-bulge.

Clarifying the origin of this discrepancy is not only essential for accurate mass modeling but also has important implications for dark matter studies. In particular, the central few kpc region of the Milky Way is sensitive to the inner shape of the dark matter distribution—a central issue in the long-standing core–cusp problem \citep[][for a review]{deBlok2010}. Among various alternatives to standard cold dark matter (CDM), ultra-light dark matter (ULDM) models predict the formation of solitonic cores—self-gravitating, cored density structures that arise from quantum pressure on de~Broglie scales \citep[][]{Schive+2014,Hui2021,Ferreira2021}. For particle masses in the range $m \sim 10^{-22}$–$10^{-21}$eV, such soliton cores could produce a detectable bump in the rotation curve or influence the kinematics of the nuclear gas disk at $R \sim 100$–$500\mathrm{pc}$ \citep[][]{Bar+2018,Li+2020}. Therefore, disentangling bar-induced non-circular motions from true mass signatures is crucial for using inner rotation curves to place meaningful constraints on dark matter models such as ULDM.

Although it has long been recognized that non-circular motions driven by the Galactic bar can bias estimates of the circular speed in the inner Milky Way \citep[e.g.,][]{Binney+1991,Fux1999,Sormani+2015a,Chemin+2015,Li+2022,Hunter+2024}, gas-based rotation curves, derived from the terminal-velocity method, are still frequently used as if they directly trace the true mass distribution. This study does not aim to introduce a new concept, but rather to provide a clear, Milky Way-specific demonstration that the steep inner rise in the gas-based circular speed curve can be reproduced solely by bar-induced streaming motions.
We conduct three-dimensional hydrodynamic simulations of the interstellar medium in a fixed, observationally motivated gravitational potential. The potential includes contributions from the stellar bar, NSD and NSC, as well as the outer stellar disk and dark matter halo, based on constraints from recent stellar dynamical models \citep[][]{Portail+2017,Sormani+2020nsd,Eilers+2019}. By generating synthetic longitude-velocity ($l$-$v$) diagrams and extracting terminal velocities from the simulated gas, we assess whether the steep rise observed in gas-based rotation curves can be reproduced solely by bar-induced non-circular streaming motions.

While recent studies have incorporated non-equilibrium chemical networks and radiative cooling to capture the detailed thermodynamics of the ISM in the inner Milky Way \citep[][]{Tress+2020,Sormani+2020}, our aim is not to resolve small-scale thermal processes. Instead, we adopt a widely used tabulated cooling function that captures the essential thermal behavior of a multi-phase ISM while remaining computationally efficient. Compared to \citet{Li+2022} and \citet{Hunter+2024}, who adopted a similar gravitational potential but assumed an isothermal gas without star formation or feedback, our model represents an intermediate level of physical realism that balances accuracy and computational feasibility. Notably, neither of these studies explicitly investigated the circular speed curve, whereas our analysis focuses on whether bar-driven gas dynamics alone can account for the steep rise inferred from the terminal-velocity method.

\section{Galaxy Models and Numerical Methods}
\label{sec:model}

We perform three-dimensional hydrodynamic simulations using the $N$-body/smoothed particle hydrodynamics (SPH) code \texttt{ASURA-3} \citep{Saitoh+2008, Saitoh2017}, which incorporates radiative cooling and heating, star formation, and stellar feedback. Hydrodynamics are solved using the density-independent SPH (DISPH) method \citep{SaitohMakino2013}, which accurately captures contact discontinuities and fluid instabilities. The self-gravity of the gas is neglected in our simulations, as our goal is to examine how gas responds to a barred external gravitational potential dominated by the stellar and dark matter components.

The gas is subject to radiative cooling and heating. The cooling function $\Lambda(T, f_{\rm H_2}, G_0)$ spans $20~\mathrm{K} \le T \le 10^8~\mathrm{K}$ and depends on gas temperature $T$, molecular hydrogen fraction $f_{\rm H_2}$ \citep{GnedinKravtsov2011}, and the far-ultraviolet (FUV) radiation field $G_0$, which is fixed at $G_0 = 1.7$ \citep{Habing1968}. Heating includes photoelectric effects \citep{Wolfire+1995} and stellar feedback. Star formation occurs stochastically in dense ($n_{\rm H} > 100~\mathrm{cm}^{-3}$), cold ($T < 100~\mathrm{K}$), and converging ($\nabla \cdot \vec{v} < 0$) gas regions \citep{Saitoh+2008}. Supernovae release $10^{51}~\mathrm{erg}$ of thermal energy per event \citep{SaitohMakino2009}, and ionizing feedback from H\,II regions is modeled using a Strömgren-sphere approximation \citep{Baba+2017}.

Our gravitational potential model is primarily based on that of \citet{Hunter+2024}, which incorporates a realistic combination of bar, disk, and nuclear components constrained by recent observations. We follow the same general framework, with minor adjustments to the stellar disk parameters.
The total gravitational potential is defined as
\begin{align}
\Phi_{\mathrm{tot}}(R,\phi,z) &= \Phi_{\mathrm{bar}}(R,\phi,z) + \Phi_{\mathrm{disk}}(R,z) + \Phi_{\mathrm{DM}}(r) \notag \\
&\quad + \Phi_{\mathrm{NSD}}(R,z) + \Phi_{\mathrm{NSC}}(R,z) + \Phi_{\mathrm{SMBH}}(r),
\end{align}
and is assumed to be static in the frame co-rotating with the bar. Here, $r = \sqrt{R^2 + z^2}$ denotes the spherical radius used in the dark matter (DM) halo and supermassive black hole (SMBH) potentials.
The bar potential $\Phi_{\mathrm{bar}}$ is based on the made-to-measure model of \citet{Portail+2017}, analytically approximated by \citet{Sormani+2022nsd}, and includes vertically extended bulge and long-bar components. The bar rotates at a pattern speed of $\Omega_{\rm b} = 37.5~\mathrm{km~s^{-1}~kpc^{-1}}$ \citep{ClarkeGerhard2022, Li+2022}, and its non-axisymmetric components are gradually introduced over the first 150 Myr to avoid numerical artifacts.
The stellar disk potential $\Phi_{\mathrm{disk}}$ consists of thin and thick components, each following a generalized exponential profile with an inner truncation \citep{McMillan2017}:
\begin{equation}
\rho_{\rm disk}(R,z) = \frac{\Sigma_0}{2h} \exp\left(-\frac{R}{R_{\rm d}} - \frac{R_{\rm cut}}{R} - \frac{|z|}{h} \right).
\end{equation}
For the thin disk, we adopt $\Sigma_0 = 2.2 \times 10^9~M_\odot~\mathrm{kpc}^{-2}$, $R_{\rm d} = 2.6~\mathrm{kpc}$, $R_{\rm cut} = 3.5~\mathrm{kpc}$, and $h = 0.3~\mathrm{kpc}$; for the thick disk, $\Sigma_0 = 2.0 \times 10^9~M_\odot~\mathrm{kpc}^{-2}$, $R_{\rm d} = 2.0~\mathrm{kpc}$, and $h = 0.9~\mathrm{kpc}$. These parameters are slightly tuned to match the stellar rotation curves derived from red giant branch stars \citep[e.g.,][]{Eilers+2019, Zhou+2023}.
The DM halo potential $\Phi_{\mathrm{DM}}$ follows a spherical Einasto profile \citep{Einasto1969}:
\begin{equation}
\rho_{\mathrm{DM}}(r) = \rho_0 \exp\left[-\left(\frac{r}{r_{\mathrm{cut}}}\right)^\eta\right],
\end{equation}
with a total halo mass of $M_{\mathrm{h}} = 1.1 \times 10^{12}~M_\odot$, a scale radius of $r_{\mathrm{cut}} = 20.5~\mathrm{kpc}$, and a shape index of $n = 4.5$ \citep{Eilers+2019}.

The nuclear stellar disk (NSD) potential $\Phi_{\mathrm{NSD}}$ is based on the axisymmetric model of \citet{Sormani+2020nsd}, in which the stellar density is expressed as a sum of two oblate exponential components:
\begin{equation}
\rho_{\mathrm{NSD}}(R,z) = \rho_1 \exp\left[-\left(\frac{a}{R_1}\right)^{n_1}\right] + \rho_2 \exp\left[-\left(\frac{a}{R_2}\right)^{n_2}\right],
\end{equation}
where $a = \sqrt{R^2 + z^2/q^2}$ with $q = 0.37$, and the parameters are $R_1 = 5.06~\mathrm{pc}$, $n_1 = 0.72$, $R_2 = 24.6~\mathrm{pc}$, $n_2 = 0.79$, and $\rho_1 / \rho_2 = 1.311$.
We also include a nuclear star cluster (NSC) component, $\Phi_{\mathrm{NSC}}$, following the flattened $\gamma$-model of \citet{Chatzopoulos+2015}, which reproduces the observed stellar density distribution in the central few parsecs. The NSC is modeled as an oblate, spherically deprojected profile with flattening $q = 0.73$, total mass $M_{\mathrm{NSC}} = 6.1 \times 10^7~M_\odot$, scale radius $a_0 = 5.9~\mathrm{pc}$, and inner slope $\gamma = 0.71$. The corresponding density distribution is
\begin{equation}
\rho_{\mathrm{NSC}}(R, z) = \frac{(3 - \gamma) M_{\mathrm{NSC}}}{4 \pi q} \frac{a_0}{a^\gamma (a + a_0)^{4 - \gamma}}.
\end{equation}
Finally, the SMBH at the Galactic center is modeled by a Plummer potential, $\Phi_{\mathrm{SMBH}}$, with a mass of $M_{\mathrm{BH}} = 4 \times 10^6~M_\odot$ and a softening length of $\epsilon_{\mathrm{BH}} = 1~\mathrm{pc}$ to avoid numerical divergence.

The gravitational potential and corresponding accelerations are computed from the density distributions of each component using the \texttt{AGAMA} code \citep[][]{Vasiliev2019}. The circular speed curve derived from the ``axisymmetric'' ($m=0$) component of the potential—hereafter referred to as the ``true circular speed curve''—is shown in Figure~\ref{fig:rc_comp}. It agrees well with those inferred from stellar kinematics across the Galactic disk \citep[e.g.,][]{Eilers+2019, Zhou+2023}. The numerical values of the circular speed curves shown in Figure~\ref{fig:rc_comp} are provided in machine-readable format as an online supplementary file (see Appendix~\ref{sec:appendix}).

The initial gas distribution is axisymmetric and exponential in both radial and vertical directions:
\begin{equation}
\rho_{\rm gas}(R,z) = \frac{\Sigma_0}{4 z_{\rm d}} \exp\left(-\frac{R}{R_{\rm d}} - \frac{|z|}{z_{\rm d}}\right),
\end{equation}
where $\Sigma_0 = 25~M_\odot~\mathrm{pc}^{-2}$ is the central surface density, $R_{\rm d} = 11$ kpc is the radial scale length, and $z_{\rm d} = 10$ pc is the vertical scale height. SPH particles are initially distributed within a cylindrical region extending to $R = 10$ kpc. All SPH particles are assigned a metallicity of [Fe/H]=0 and an initial temperature of $T = 10^4$ K. The initial velocity field assumes purely circular rotation based on the axisymmetric component of the gravitational potential. In addition, a random velocity component is added, drawn from a Gaussian distribution with a velocity dispersion of $10~\mathrm{km~s^{-1}}$.
The initial mass of each SPH particle is $3000~M_\odot$.

\section{Bias in Gas-based Circular Speed Curves due to Bar-induced Streaming}

\begin{figure*}
\begin{center}
\includegraphics[width=0.95\textwidth]{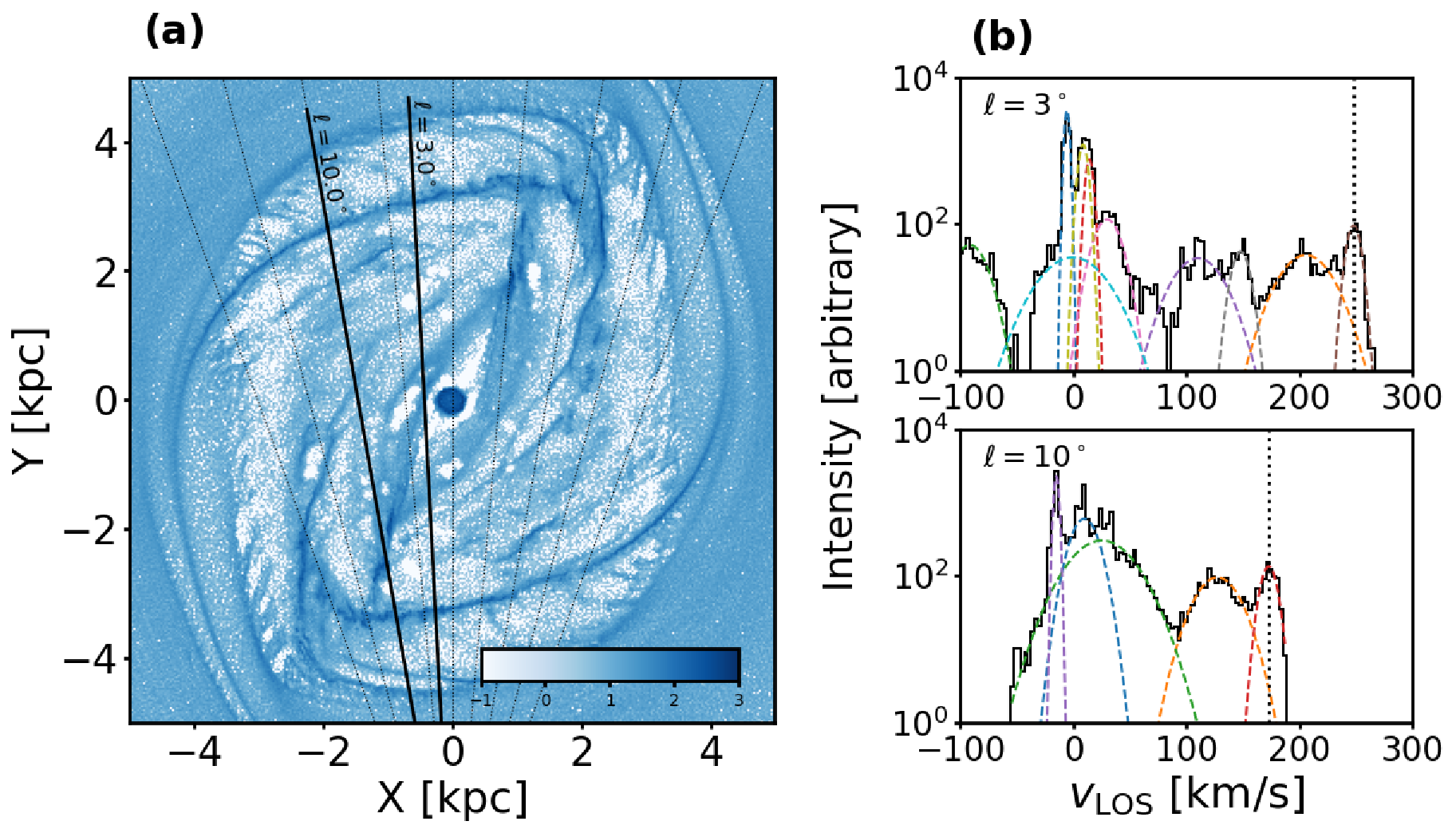}
\end{center}
\caption{
(a) Face-on view of the gas surface density at $t = 300~\mathrm{Myr}$ from our hydrodynamic simulation. The Galactic bar is inclined by $\phi_{\rm bar} = 25^\circ$ clockwise from the Sun–Galactic center line. The observer (LSR) is located at $(x,y) = (0, -8.3)~\mathrm{kpc}$ and moves with velocity $(v_x, v_y) = (-235, 0)~\mathrm{km~s^{-1}}$ in the simulation frame. The gas surface density is shown in logarithmic scale in units of $\mathrm{M}_\odot\,\mathrm{pc}^{-2}$. Characteristic features such as a $\sim 200$~pc nuclear gas disk, leading-edge dust lanes, and trailing spiral arms are visible.
(b) LOS velocity distributions (histograms; solid lines) for cold gas around Galactic longitudes $l = 3.0^\circ$ and $10.0^\circ$, integrated over $\pm 0.5^\circ$ in longitude. Gaussian mixture model (GMM) fits are overplotted as dashed curves. For each direction, the terminal velocity $V_{\rm TP}(\ell)$ is indicated by a vertical dashed line corresponding to the mean of the Gaussian component with the largest absolute velocity.
}
\label{fig:snapshot}
\end{figure*}

We begin by examining the gas distribution in our simulation to provide context for the kinematic analysis that follows. Figure~\ref{fig:snapshot}(a) presents a face-on view of the gas surface density at $t = 300~\mathrm{Myr}$. The Galactic bar is inclined $\phi_{\rm bar} = 25^\circ$ clockwise from the Sun–Galactic center line, with the Sun (LSR) located at $(x,y) = (0, -8.3)~\mathrm{kpc}$ and moving with velocity $(v_x, v_y) = (-235, 0)~\mathrm{km~s^{-1}}$ in the simulation frame. These values correspond to a Galactocentric distance $R_0 = 8.3~\mathrm{kpc}$ and circular speed $V_0 = 235~\mathrm{km~s^{-1}}$ \citep[][]{Bland-HawthornGerhard2016,HuntVasiliev2025}. The gas responds strongly to the non-axisymmetric bar potential, forming characteristic structures such as a $\sim 200$-$300$~pc nuclear gas disk and leading-edge offset dust lanes extending to $R \sim 3~\mathrm{kpc}$. Beyond these dust lanes, trailing gaseous spiral arms emerge in the outer disk. These features are consistent with previous simulations of barred galaxies \citep[e.g.,][]{Athanassoula1992b,Sormani+2020,Li+2022,Hunter+2024,Dura-Camacho+2024}.

\subsection{Overestimation of Circular Speed and Enclosed Mass Profiles}
\label{sec:bar_rc}

\begin{figure*}
\begin{center}
\includegraphics[width=0.95\textwidth]{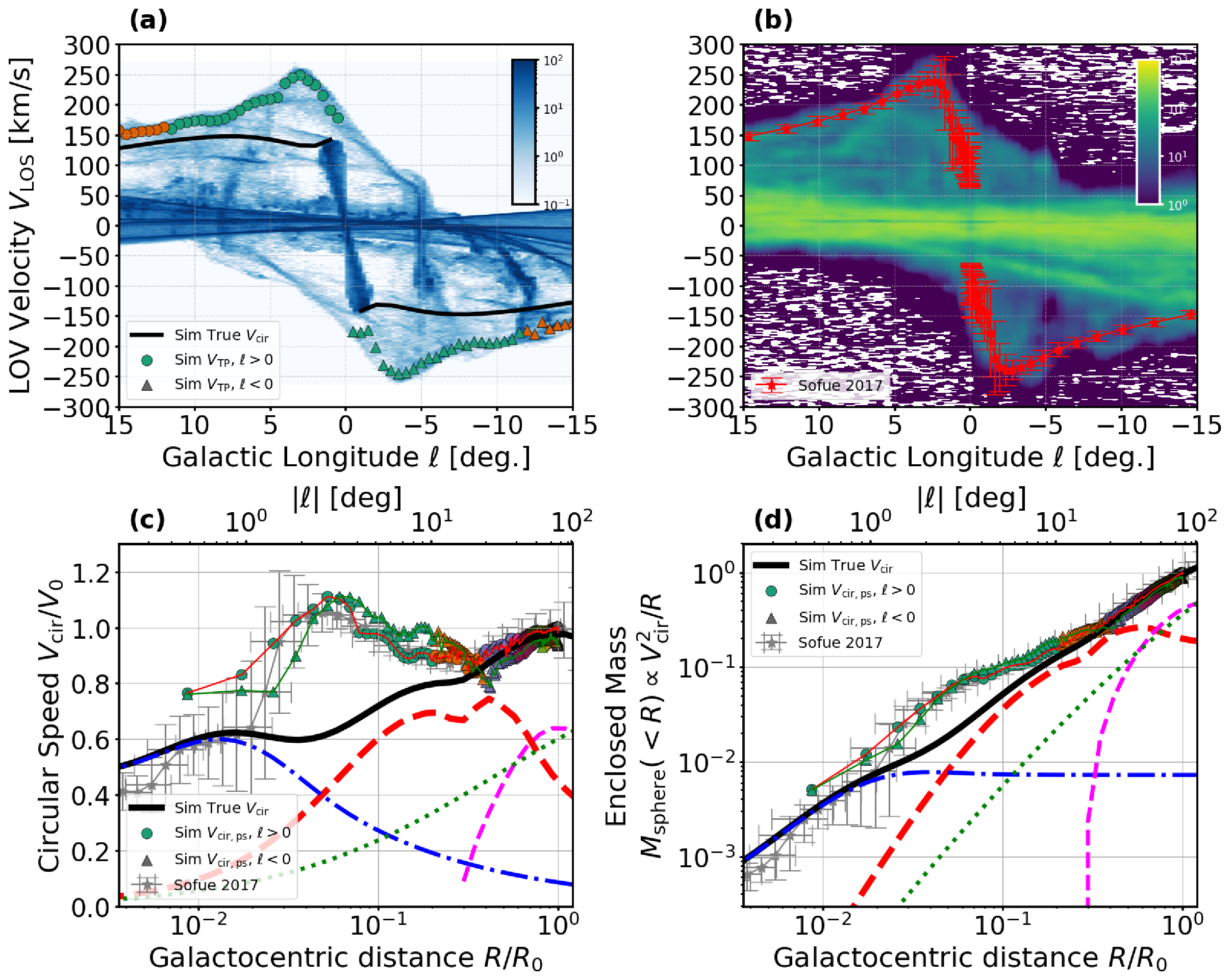}
\end{center}
\caption{
Kinematic analysis of the inner Milky Way at $t = 300~\mathrm{Myr}$ based on the hydrodynamic simulation.
\textbf{(a)} Simulated longitude–velocity ($l$–$v$) diagram for cold gas particles. 
Terminal velocities $V_{\rm TP}$, extracted using a Gaussian mixture model (GMM), are overplotted: circles for $l > 0^\circ$ and triangles for $l < 0^\circ$. 
The solid black line shows the LOS velocity corresponding to the true circular speed.
\textbf{(b)} Observed HI $l$–$v$ diagram from the Leiden/Argentine/Bonn Survey \citep{Kalberla+2005}, shown for comparison.
\textbf{(c)} Pseudo circular speed profile $V_{\mathrm{cir,ps}}(R)$ obtained from the simulated $V_{\rm TP}$ values using the standard tangent-point method under the assumption of axisymmetry. 
The gray points with error bars represent the observed pseudo circular speed profile from \citet{Sofue2017}. 
The black thick solid line shows the true circular speed. Colored lines indicate contributions from individual mass components (bar, disk, dark matter, NSD+NSC+SMBH), as in Figure~\ref{fig:rc_comp}.
At $R/R_0 \approx 0.05$ (i.e. $R \approx 0.4$ kpc), the pseudo circular speed exceeds the true value by nearly a factor of two.
\textbf{(d)} Enclosed mass profiles estimated under the spherical approximation, $M(<R) = R V_{\mathrm{cir}}^2/G$, using both the pseudo and true circular speed curves. Colored curves correspond to the mass components shown in panel \textbf{(c)}. 
The pseudo circular speed profiles (from both simulation and observation) significantly overestimate the enclosed mass in the inner Milky Way, by up to a factor of four around $R/R_0 \approx 0.05$.
}
\label{fig:lvmap}
\end{figure*}

Figure~\ref{fig:lvmap}(a) shows the simulated longitude–velocity ($l$–$v$) diagram for cold gas ($T < 10^4$~K). Parallelogram-shaped envelopes, tilted ridges, and multi-peaked features—commonly observed in the Milky Way—are well reproduced in the simulation. For comparison, the observed H\,\textsc{I} 21~cm $l$–$v$ map from the Leiden/Argentine/Bonn Survey \citep{Kalberla+2005} is shown in Figure~\ref{fig:lvmap}(b).

To extract terminal velocities, we apply a Gaussian mixture model (GMM) to the LOS velocity ($v_{\rm LOS}$) distribution at each Galactic longitude ($l$). The terminal velocity $V_{\rm TP}$ is defined as the mean of the Gaussian component with the largest absolute velocity. This approach avoids ad hoc thresholds and captures coherent motions near tangent points. Figure~\ref{fig:snapshot}(b) illustrates two representative examples at $l = 3^\circ$ and $10^\circ$, showing the velocity histograms and their corresponding GMM fits, with the component yielding $V_{\rm TP}$ indicated. The resulting $V_{\rm TP}$ values are overplotted on the simulated $l$–$v$ diagram in Figure~\ref{fig:lvmap}(a), using circle markers for positive longitudes ($l > 0^\circ$) and triangle markers for negative longitudes ($l < 0^\circ$). The overplotted $V_{\rm TP}$ values closely trace the envelope of the simulated $l$–$v$ diagram, indicating that the GMM-based method reliably identifies the terminal velocity features. For reference, the line-of-sight projection of the true circular speed—computed from the axisymmetric ($m=0$) component of the gravitational potential—is shown as a solid black line. Within $|l| \lesssim 15^\circ$, both the gas emission envelope and the extracted $V_{\rm TP}$ exceed the projected true circular speed, indicating significant bar-induced streaming motions that lead to an overestimation of the circular speed.

Using the standard tangent-point relation \citep[e.g.][]{Sofue2017},
\begin{equation}
\label{eq:TP}
    \frac{V_{\mathrm{cir,ps}}(R)}{V_0} = \sin l + \frac{V_{\rm TP}(l)}{V_0}, \quad \sin l = \frac{R}{R_0},
\end{equation}
we derive a ``pseudo'' circular speed, $V_{\mathrm{cir,ps}}(R)$, under the assumption of purely circular motion. The resulting profile is shown in Figure~\ref{fig:lvmap}(c), along with the observed pseudo circular speed curve derived from CO and H\,I data using the same method \citep[][gray asterisks with error bars]{Sofue2017}. The simulated profile exhibits a steep rise in the inner Milky Way ($R/R_0 \lesssim 0.3$), closely resembling the observational result. However, both curves lie significantly above the true circular speed (solid black line), especially at $R/R_0 \approx 0.05$ (i.e. $R \approx 0.4$ kpc), where $V_{\mathrm{cir,ps}}$ overestimates the true circular speed by nearly a factor of 2.

Figure~\ref{fig:lvmap}(d) shows the enclosed mass profile estimated under a spherical approximation using $M_{\mathrm{sphere}}(<R) = R\,V_{\mathrm{cir}}^2/G$. We compute this profile for both the simulated and observed pseudo circular speed curves (from Figure~\ref{fig:lvmap}c), and also for the true circular speed (solid black line). In the radial range $0.03 \lesssim R/R_0 \lesssim 0.3$ (i.e. $0.25 \lesssim R \lesssim 2.5$ kpc), the enclosed mass inferred from the pseudo circular speeds significantly exceeds the true enclosed mass—by up to a factor of 4 at $R/R_0 \approx 0.05$. This overestimation arises from the breakdown of the tangent-point method’s assumptions, particularly the neglect of non-circular streaming motions induced by the bar. Our finding is consistent with previous suggestions based on hydrodynamic simulations that bar-induced streaming motions can bias the enclosed mass estimates from gas kinematics in the inner Milky Way \citep[][]{KodaWada2002, Chemin+2015}.

\subsection{Biases in Terminal Velocities Induced by Bar-driven Orbits}
\label{sec:orbit}

To clarify the physical origin of the overestimated circular velocities and enclosed mass profiles identified in Section~\ref{sec:bar_rc}, we now examine how bar-driven orbital structures systematically bias terminal velocity measurements derived from gas kinematics. In barred galaxies, both stars and gas tend to follow elongated, non-circular orbits aligned with the bar. The dominant family in the inner region is the $x_1$ orbit family, which supports the bar’s structure and provides a useful dynamical framework for understanding non-circular motions \citep[][]{ContopoulosGrosbol1989}. While real gas dynamics are modified by dissipation, shocks, and pressure forces \citep[e.g.,][]{Wada1994, Sormani+2015b}, the large-scale structure of the flow is often governed by the $x_1$ orbital backbone \citep[e.g.,][]{Binney+1991, Athanassoula1992a}

\begin{figure*}
\begin{center}
\includegraphics[width=0.98\textwidth]{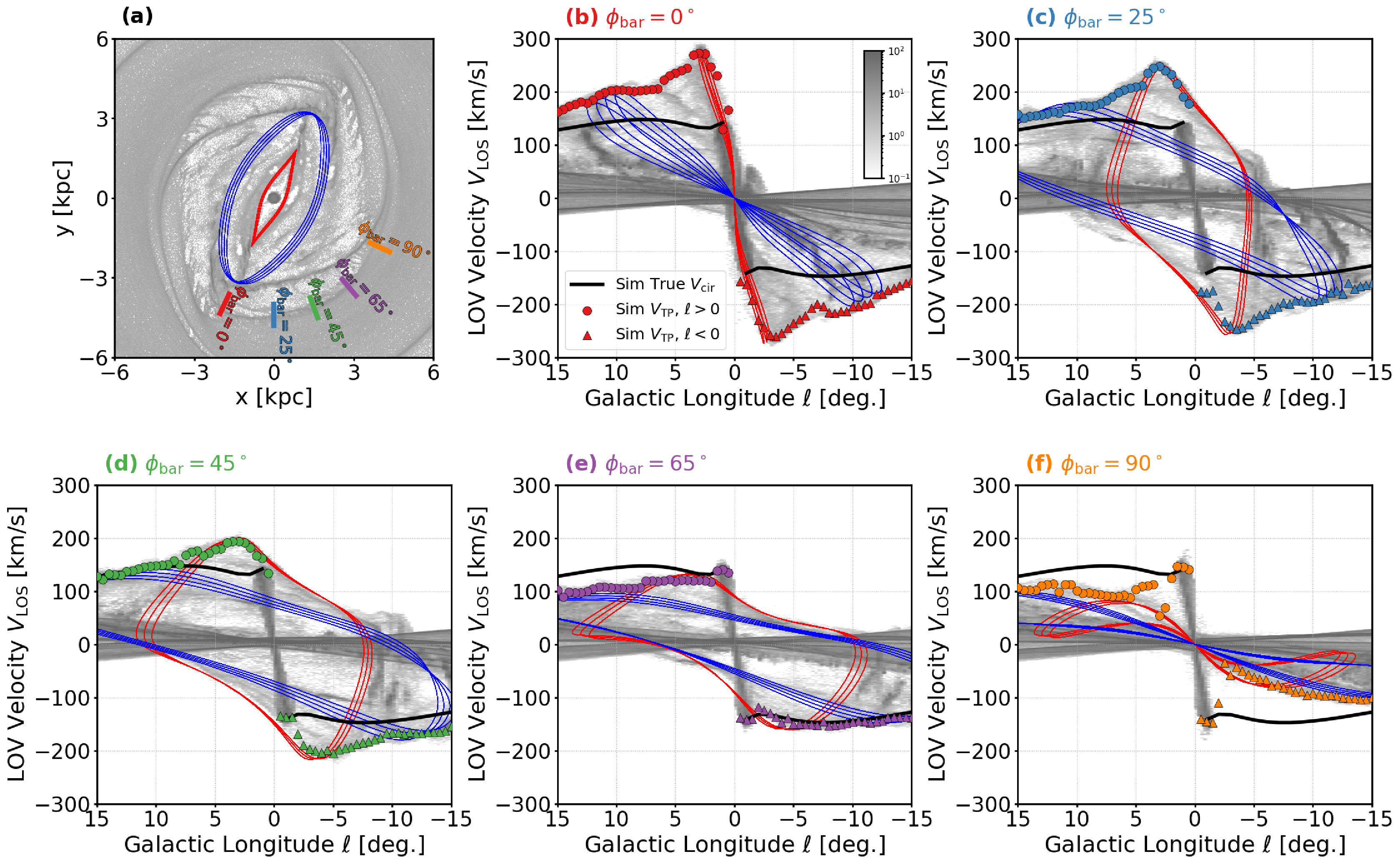}
\end{center}
\caption{
Effect of bar viewing angle on simulated longitude–velocity ($l$–$v$) diagrams.  
(a) Face-on view of the gas surface density at $t = 300~\mathrm{Myr}$, with arrows indicating the lines of sight for different bar viewing angles: $\phi_{\rm bar} = 0^\circ$, $25^\circ$, $45^\circ$, $65^\circ$, and $90^\circ$ measured clockwise from the bar's major axis.  
(b)--(f) Simulated $l$–$v$ diagrams corresponding to each $\phi_{\rm bar}$ value. In each panel, the solid black curve shows the line-of-sight projection of the true circular speed. 
Filled circles (for $\ell > 0^\circ$) and triangles (for $\ell < 0^\circ$) indicate terminal velocities extracted using GMM fits to the LOS velocity distributions.
}
\label{fig:bar_orbit_angle}
\end{figure*}

Figure~\ref{fig:bar_orbit_angle}(a) shows representative $x_1$ orbits in our adopted barred potential, with inner and outer trajectories highlighted in red and blue, respectively. Because these orbits are elongated along the bar, the projected LOS velocities strongly depend on the viewing angle $\phi_{\rm bar}$. When the bar is viewed nearly end-on (i.e., $\phi_{\rm bar} \sim 0^\circ$), the line of sight intersects the pericenters, where gas moves rapidly toward or away from the observer, leading to an overestimate of the true circular speed defined by the axisymmetric component of the potential. In contrast, a side-on view ($\phi_{\rm bar} \sim 90^\circ$) intersects the slower apocentric regions, yielding an underestimate \citep[][]{KodaWada2002, Chemin+2015}.

To visualize this projection effect, we generated synthetic longitude–velocity ($l$–$v$) diagrams from our simulation for various $\phi_{\rm bar}$ (Figures~\ref{fig:bar_orbit_angle}b–f). In each panel, we overplot the projected $x_1$ orbits. These trace characteristic parallelogram-like structures in $l$–$v$ space, whose outer edges closely follow the terminal velocity envelope of the gas. This agreement demonstrates that the velocity structure of the inner gas is largely shaped by $x_1$-like streaming motions \citep[e.g.,][]{EnglmaierGerhard1999, Sormani+2015a}.

\begin{figure}
\begin{center}
\includegraphics[width=0.48\textwidth]{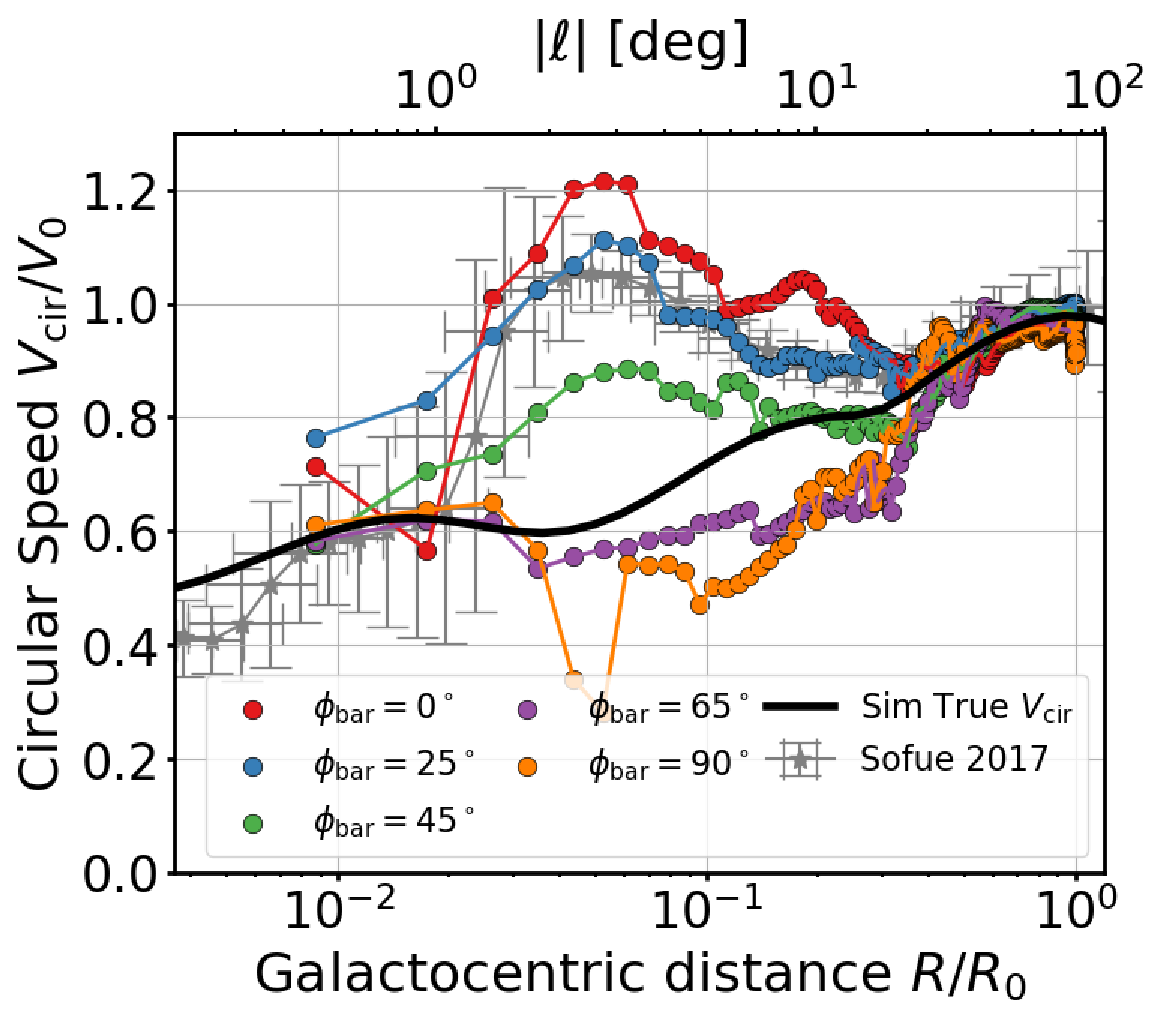}
\end{center}
\caption{
Pseudo circular speed curves $V_{\mathrm{cir,ps}}(R)$ derived from terminal velocities at different bar viewing angles $\phi_{\rm bar}$, using the standard tangent-point method assuming purely circular motion. Each colored curve corresponds to a different $\phi_{\rm bar}$ from the simulated $l$–$v$ diagrams shown in Figure~\ref{fig:bar_orbit_angle}. For clarity, only terminal velocities at positive longitudes ($l > 0^\circ$) are used to compute these curves. While the underlying bar potential is symmetric, the projected gas kinematics are not strictly symmetric between $l > 0^\circ$ and $l < 0^\circ$ due to the bar’s asymmetry in viewing geometry. Nevertheless, the qualitative behavior of the pseudo circular speed curves remains similar between the two sides. The gray points with error bars show the observed pseudo circular speed curve from \citet{Sofue2017}. The solid black line indicates the true circular speed curve computed from the axisymmetric ($m=0$) component of the gravitational potential in the simulation.
}
\label{fig:rc_barangle}
\end{figure}

The extent of this envelope varies systematically with $\phi_{\rm bar}$: for small angles (e.g., $\phi_{\rm bar} = 0^\circ$), terminal velocities are higher because of alignment with fast pericentric motions; for large angles (e.g., $\phi_{\rm bar} = 90^\circ$), the envelope narrows due to lower projected speeds. To quantify the resulting bias in inferred mass profiles, we applied the tangent-point method (equation~\ref{eq:TP}) to extract pseudo circular speed curves $V_{\mathrm{cir,ps}}(R)$ for each angle (Figure~\ref{fig:rc_barangle}). The results show that for $\phi_{\rm bar} \lesssim 45^\circ$, $V_{\mathrm{cir,ps}}$ exceeds the true circular speed in the inner Galaxy ($R/R_0 \lesssim 0.4$), mimicking a steep central rise. For $\phi_{\rm bar} \gtrsim 60^\circ$, the opposite trend appears.
These findings confirm that terminal velocities—and hence inferred circular speeds—are significantly affected by the viewing geometry of the bar. Even when the same analysis method is applied, projection effects can produce large systematic biases. Notably, the pseudo circular speed curve at $\phi_{\rm bar} = 25^\circ$ closely reproduces the gas-based profile reported by \citet{Sofue2017}, suggesting that the apparent central mass excess in such curves arises from bar-induced streaming motions rather than from a true increase in central mass.

\section{Discussion and Summary}
\label{sec:Discussion}

Our simulations demonstrate that the standard terminal-velocity method, when applied to gas kinematics in the inner Milky Way, systematically overestimates the ``pseudo'' circular speed—the circular speed inferred from the maximum observed LOS gas velocity under the assumption of axisymmetry \citep[][and references therein]{Sofue2017}. This bias is most prominent within $0.1 \lesssim R \lesssim 4~\mathrm{kpc}$, where the dynamical influence of the Galactic bar dominates and gas motions deviate substantially from circular rotation. In particular, because the Sun is located near the major axis of the bar, the projected LOS velocities are strongly affected by bar-induced streaming motions. As a result, the pseudo circular speed curve can exhibit steep inner rises or artificial bumps that do not reflect the true underlying mass distribution.

Our key finding is that such features can be reproduced without invoking additional central mass components. By simulating gas flow in a realistic barred gravitational potential—constructed from stellar dynamical models calibrated to observational data \citep[][]{Portail+2017, Sormani+2022agama, Sormani+2020nsd, Hunter+2024}—we recover the steep inner rise in the pseudo circular speed. This validates earlier theoretical studies that reported similar effects using idealized bar models \citep[][]{KodaWada2002, Chemin+2015}, and provides a Milky Way-specific, dynamically consistent explanation. The implication is clear: steep inner features in gas-based rotation curves can largely arise from bar-induced non-circular motions, rather than from a genuine central mass concentration.

This has significant consequences for how inner mass distributions are interpreted. For instance, \citet{Sofue+2009} attributed the steep rise in the pseudo circular speed curve within $0.1 \lesssim R \lesssim 4$ kpc to a massive classical bulge with $M_{\mathrm{bulge}} \approx 1.8 \times 10^{10}\, M_\odot$, based on a de Vaucouleurs profile fit. Later, \citet{Sofue2013} proposed an exponential bulge model for the same region, yielding a somewhat lower mass of $8.4 \times 10^9\, M_\odot$. However, our results indicate that the terminal-velocity-based pseudo circular speed in this region is significantly biased by bar-driven streaming motions. Using these data to estimate mass under the assumption of circular motion leads to a substantial overestimation. If our interpretation holds, the true enclosed mass at $R \sim 1$ kpc may be overestimated by a factor of 2–4, implying a more realistic mass of a few $\times\, 10^9\, M_\odot$. This is consistent with the findings of \citet{Shen+2010}, who showed through comparisons with $N$-body models that the bulge can be explained without invoking a massive classical component, constraining its mass to less than $\sim$8\% of the disk mass.

The overestimation of the inner mass profile also affects constraints on dark matter models. In particular, ULDM models predict the formation of solitonic cores—self-gravitating, cored density structures stabilized by quantum pressure on de Broglie scales \citep[][for reviews]{Hui2021, Ferreira2021}. For a Milky Way-sized halo ($M_{\mathrm{h}} \sim 10^{12} M_\odot$) and boson mass $m \sim 10^{-22}$ eV, the expected soliton core has a mass of $\sim 1.4 \times 10^9 M_\odot$ and radius of $\sim 160$ pc \citep[][]{Schive+2014, Bar+2018}. Observed steep features in gas-based rotation curves within $100 \lesssim R \lesssim 500$ pc have been used to constrain such cores. However, our results caution against interpreting these features as evidence of soliton cores without correcting for bar-induced non-circular motions, as doing so could lead to biased estimates of both the dark matter distribution and the ULDM particle mass.

A more robust approach involves stellar dynamical modeling. Unlike gas, stars behave as a collisionless system and are not subject to shocks, pressure gradients, or radiative losses. Although individual stellar orbits can be non-circular, their statistical properties can be used to infer the underlying gravitational potential using techniques such as Jeans analysis or distribution function modeling. Recent studies have successfully applied these methods to the inner Milky Way, including the bar, NSD, and NSC regions \citep[][]{Sormani+2020nsd, Sormani+2022nsd}, and to test the presence of soliton-like dark matter cores \citep[][]{Toguz+2022}.

Finally, we emphasize that the limitations and biases introduced by assuming circular motion in non-axisymmetric regions have long been recognized \citep[e.g.,][]{Binney+1991,Fux1999,Sormani+2015c,Chemin+2015,Hunter+2024}. Nevertheless, the terminal-velocity curve—particularly that of \citet{Sofue2017}—continues to be widely used as a proxy for the true circular speed, especially in dark matter studies. In such contexts, the steep rise observed in gas-derived rotation curves is sometimes interpreted as evidence for a massive central concentration, including exotic scenarios such as ultralight dark matter solitons. Our results demonstrate that this feature can arise naturally from bar-driven streaming motions alone, without invoking a massive classical bulge or non-standard dark matter distributions. This highlights the importance of using dynamical models that explicitly incorporate non-circular motions when deriving mass estimates in the inner Milky Way.

\begin{acknowledgments}
We are grateful to Takayuki Saitoh for providing and supporting the {\tt ASURA-3} code. We also thank Keiichi Wada for encouraging us to publish this work. 
We sincerely thank the anonymous referee for their thoughtful and constructive comments, which helped improve the clarity and context of this paper.
Calculations, numerical analyses, and visualization were conducted on Cray XD2000 (ATERUI-III), and computers at the Center for Computational Astrophysics, National Astronomical Observatory of Japan (CfCA/NAOJ). This research was supported by the Japan Society for the Promotion of Science (JSPS) under Grant Numbers JP21K03633, JP21H00054, JP22H01259, JP24K07095, and JP25H00394.
\end{acknowledgments}

\appendix
\section{Circular Speed Curve Data}
\label{sec:appendix}

The file \texttt{mw\_circular\_speed\_JB.dat} provides the machine-readable numerical values for the circular speed curves shown in Figure~\ref{fig:rc_comp}. These include the contributions from individual mass components: the axisymmetric ($m=0$) part of the bar, the stellar disk (including both thin and thick disks), the dark matter halo, and the central components (comprising the nuclear stellar disk, nuclear star cluster, and central black hole). This file is available as online supplementary material.

We note that, since this study focuses primarily on the inner Milky Way, the circular speed curves beyond the solar radius (i.e., in the outer disk and halo) have not been extensively validated against observational constraints.


\end{document}